Full Articles

# Angular optimization for cancer identification with circularly polarized light

Nozomi Nishizawa*,1 | Bassam Al-Qadi2 | Takahiro Kuchimaru3

1 Laboratory for Future Interdisciplinary Research and Technology, Tokyo Institute of Technology, Yokohama 226-8503, Japan.

2 College of Engineering and Technology, Palestine Technical University -Kadoorie, P.O.Box: 7, Yafa Street, Tulkarm, Palestine

3 Center for Molecular Medicine, Jichi Medical University, Tochigi 329-0498, Japan.

**\* Correspondence**

Laboratory for Future Interdisciplinary Research and Technology, Tokyo Institute of Technology, Yokohama 226-8503, Japan.

Email: nishizawa.n.ab@m.titech.ac.jp

Depolarization of circularly polarized light scattered from biological tissues depends on structural changes in cell nuclei, which can provide valuable information for differentiating cancer tissues concealed in healthy tissues. In this study, we experimentally verified the possibility of cancer identification using scattering of circularly polarized light. We investigated the polarization of light scattered from a sliced biological tissue with various optical configurations. A significant difference between circular polarizations of light scattered from cancerous and healthy tissues is observed, which is sufficient to distinguish a cancerous regioin. The line-scanning experiments along a region incorporating healthy and cancerous parts indicate step-like behaviors in the degree of circular polarization corresponding to the state of tissues, whether cancerous or normal. An oblique and perpendicular incidence induces different resolutions for identifying cancerous tissues, which indicates that the optical arrangement can be selected according to the priority of resolution.

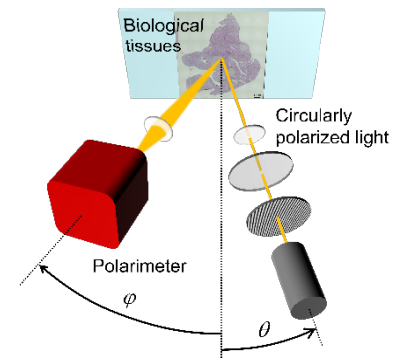

## 1 | INTRODUCTION

The multiple scattering phenomenon of polarized light provides valuable information regarding scatterers in a turbid material, based on the polarization state as well as the remnant intensity of the scattered light. When polarized light beams impinge on a biological tissue, they penetrate and propagate into the tissue and are scattered multiple times by cell nuclei, which are the main scatterers in biological tissues. Eventually, they are absorbed inside or discharged outside the tissue. Depolarization of the resultant scattered light mainly depends on the size and axial ratio of scatterers, that is, cell nuclei, as well as the frequency of scattering events associated with the density and distribution of cell nuclei.

Bickel *et al.* [1] reported that the polarization state of light scattered differentially from suspended biological scatterers can provide structural information about biological tissues. They suggested that this technique is useful for distinguishing closely related structural biological systems and for identifying subsequent time-dependent structural changes. Most of the earlier studies that were conducted to observe biological structures with the polarization state of the back-scattered light have been performed using the linear polarized (LP) light [2 - 8]. In the Mie scattering regime [9], the size of a scatterer (cell nucleus) is larger than the wavelength of incident light, and the degree of depolarization for LP light is significantly large such that the scattered light results in complete depolarization by a small number of scattering events [10, 11]. Therefore, these studies using the LP light have successfully provided scientific significance specifically for surface observations [12 -14].

In contrast, the circular polarization of light has more persistence against multiple scattering in the Mie regime. The depolarization process is caused by two randomized processes: rotations of the polarization plane and interference with the backscattered light. When the size of a scatterer is almost equivalent or smaller than the wavelength of incident light (Rayleigh regime), LP light is randomized mostly by the former process, whereas circularly polarized (CP) light is disturbed dominantly by the latter process. The strength of the resultant depolarization was almost equivalent. In the Mie regime, in which forward scattering is dominant [9], complete depolarizations of CP light require more scattering events compared with those of LP light owing to the reduction in the backward scattering [10, 11]. Therefore, scattering of CP light can provide more specific information about not only the outermost surface but also the interior of tissues, which suggests the possibility of identifying carcinoma concealed in tissues.

Meglinski *et al*. [15] pioneered the application of CP light for cancer detection by experimentally mapping the scattering properties of tissues on the Poincaré sphere. Kunnen *et al*. [16] reported that the polarization of light scattered from a tissue of human lung shows different polarization states for healthy and tumor tissues *ex vivo* using incident CP light ($\lambda = 639$ nm). They concluded that the difference in polarization is caused by the enlargement of the nucleus size due to canceration. Furthermore, they suggested that this technique would generate noninvasive diagnostic technology for early disease detection. Triggered by these reports, the tissue polarimetry technique has been widely studied to develop an optical diagnostic tool that can provide supplementary information for pathologists [17-24]. Recently, the polarimetry technique has been applied and demonstrated for grading colon cancer [25] and Alzheimer's disease [26].

We have studied spin-polarizing light-emitting diodes (spin-LEDs) that can emit fully CP light directly from the side facet and can detect CP light without applying an external magnetic field or a large electrical field [27-30]. If spin-LED devices are integrated at the tip of a biopsy probe apparatus such as an endoscope, the polarimetry technique suggested by Meglinski *et al*. can be developed from *ex vivo* observation to *in vivo* observation, which also enables observation in real time while avoiding the risk of administering a fluorescent agent. However, to develop this technique for practical use, more intensive and detailed investigations are required from both theoretical and experimental approaches.

Previously, we theoretically investigated the scattering process of CP light against cell nuclei in pseudo tissues using Monte Carlo (MC) methods based on the Mie scattering mechanism [31]. MC simulations were performed for cancerous and healthy pseudo tissues in aqueous medium containing dispersed particles with the typical sizes of cell nuclei in healthy and cancerous cells, that is, 6 μm and 11 μm, respectively. Accordingly, a distinct difference in the resultant polarization values between healthy and cancerous tissues can be obtained over a wide range of detection angles, which suggests that this technique can characterize the size of cell nuclei in biological tissues. The difference is estimated to be approximately 0.2 in the degree of circular polarization (DOCP) values.

In this study, we experimentally demonstrated the identification of cancer in sliced biological tissue with the scattering of CP light. We measured the DOCP values of scattered light in various optical angular arrangements with incident and detection angles. Line-scanning experiments were performed to demonstrate clear discrimination of the cancerous and healthy parts, which was partially published in ref. [32]. In addition, we assessed the in-plane dispersion of the detected values according to optical angular configurations.

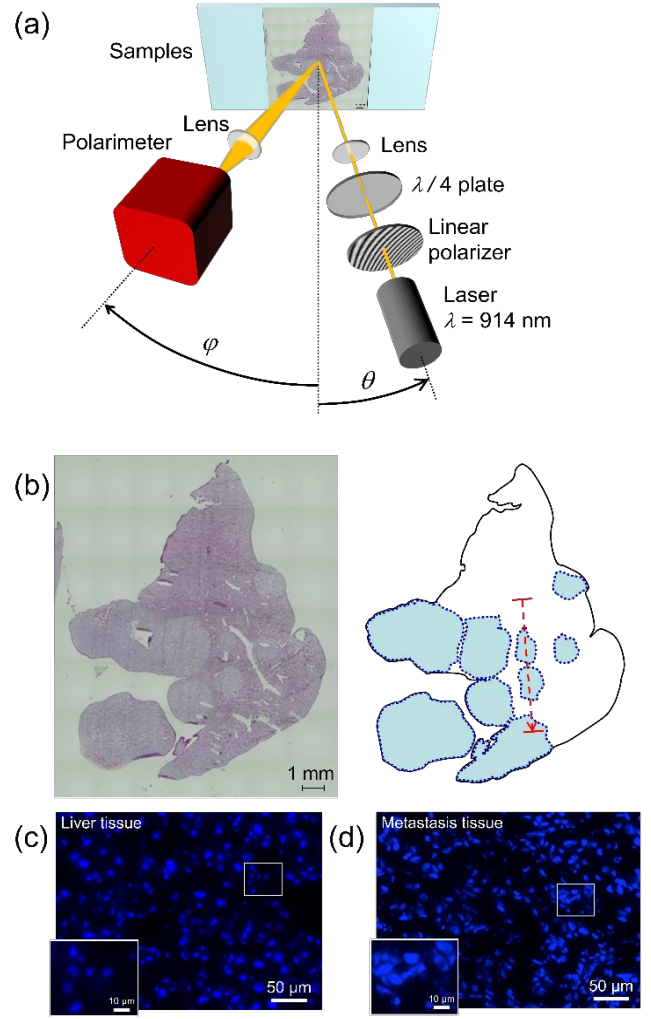

**FIGURE 1** (a) Schematic illustrations of experimental setup for measuring the DOCP value in this study. (b) (left) Micrograph of specimen and (right) a corresponding schematic map. The light-blue areas delineated by the blue dotted line represent the metastasis parts. The red dotted arrow shows the area where line-scanning experiments were performed, which is across normal and metastasis parts. Fluorescence micrographs of (c) healthy liver and (d) metastasis tissues with stained cell nuclei in the specimen. The insets show the magnified images of the region in white boxes.

## 2 | EXPERIMENTAL

Figure 1(a) shows the experimental setup used to measure the DOCP values in this study. The incident and detection angles, $\theta$ and $\varphi$, are defined as angles made by the lines connecting from a measurement point on the sample to the light source and the detector with a perpendicular line at the measurement point, respectively. The unpolarized laser beam emitted from a diode-pumped solid-state (DPSS) laser (Sanctity Laser Technology Co., Ltd., China) with a wavelength of 914 nm and a power of 100 mW was converted to right-handed CP light with 0.5 mW through an ND filter, a linear polarizer and a quarter-wave ($\lambda$/4) plate. Subsequently, it is focused by a plano-convex lens ($f = 25$ mm) at a measurement point with an incident angle $\theta$. The DOCP value of the incident CP light was + 1.0. The light scattered from the sample at an angle of $\varphi \pm 5.7°$ is collected by a plano-convex lens ($f = 30$ mm) and detected by a polarimeter with a high dynamic range (PAX1000; Thorlabs, Inc.), which consists of an optical input aperture ($\phi$ 3 mm), a rotating $\lambda/4$ plate, a fixed linear polarizer, and a photodiode. The polarization state of the scattered light is assessed by the DOCP values, which is defined by the equation, DOCP = $S_3/S_0$, where $S_0$ and $S_3$ are the Stokes polarization parameters that describe the total intensity of the scattered light and the preponderance of the right CP over the left CP, respectively. The $S_0$ values can be changed according to the difference in the reflection coefficient and scattering efficiency. However, we confirmed that the changes in circular polarization $\Delta S_3$ are considerably larger than the variations in the total intensity $\Delta S_0$; therefore, the obtained DOCP values are intrinsically derived from the circular polarization of light in all of the experiments in this study. The incident angle dependences were investigated by varying $\theta$ from 35° to 55° and fixing $\varphi$ to 0°. Conversely, the detection angle dependences were assessed by changing $\varphi$ from 35° to 55° and fixing $\theta$ to 0°.

Sliced tissue specimens of liver metastasis were prepared from a murine xenograft model with human pancreatic cancer SUIT2 cells. To establish liver metastasis in mice, human pancreatic cancer SUIT2 cells were intrasplenically injected. After 47 days of injecting cancer cells, the livers were harvested to prepare sliced specimens containing liver metastatic lesions. The livers harvested from the mice were immediately frozen in a frozen tissue matrix. The frozen liver was serially sectioned into specimens with a thickness of approximately 40 μm using a cryostat. One of the serial sections were used to measure CP scattering and the other were utilized to observe the stained cell nuclei. Figure 1(b) shows a micrograph of the specimen for measurement of CP scattering with neither staining nor any support. The liver section for the CP scattering measurements is laid on a glass plate and then tightly adhered to the glass with its moisture. The specimen with a glass plate was placed upright in the optical setup. The right illustration of Figure 1(b) is a schematic map that indicates the characteristics. The metastasis parts are shown by the light-blue area surrounded by a dotted line. The red dotted arrow denotes the linear area along which the line-scanning measurements were performed. The fluorescence micrographs of healthy liver and metastatic area with stained cell nuclei in the specimen are shown in Figures 1 (c) and (d), respectively. The liver sections were fixed with 4% paraformaldehyde for 10 min. After washing with phosphate-buffered saline, the fixed sections were stained with 10 μg/mL Hoechst 33342 for 5 min. The stained sections were sealed with cover slips and fluorescently observed with an inverted fluorescent microscope (BZ-X800, Keyence). Fluorescent images were acquired with a 40× objective lens under fixed exposure time of the CMOS camera. The cell nuclei in the healthy part are comparatively small and dispersed, and the average diameter of the nuclei is approximately 6.4 μm. In the metastasis part, relatively large and aggregated cell nuclei are observed, which indicates that the enlargement of cell nuclei is observed in metastasis parts due to canceration. The average diameter of the cell nuclei is approximately 10.8 μm.

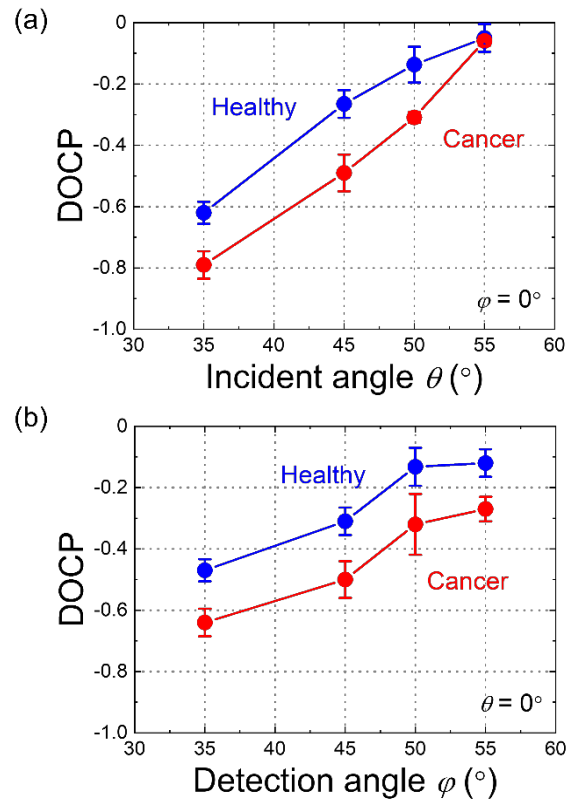

**FIGURE 2** DOCP values as a function of (a) incident angles $\theta$ with $\varphi = 0°$ and (b) detection angles $\varphi$ with $\theta = 0°$, respectively.

MC simulations for the pseudo-biological tissue with a thickness of 40 μm show that the sampling volume is approximately 1 mm × 1 mm × 40 μm, independent of the angular configurations. The sampling depth at $(\theta, \varphi) =$ (1°, 30°) is estimated to be 1.65 mm from the MC calculations for the tissue with infinite thickness [31]. Therefore, a significant portion of incident light is transparent through the back of the sample or reflected at the interface between the specimen and the glass plate. The remaining portion of light and reflected light at the back undergo multiple scattering events in the entire sample with a thickness of 40 μm and then arrive at the detector. This sampling volume includes approximately 750 nuclei on average.

## 3 | RESULTS AND DISCUSSION

Figures 2 (a) and (b) show the dependence of the DOCP value of scattered light on the incident angle $\theta$ and detection angle $\varphi$, respectively. The blue and red dots and lines represent the DOCP values measured at the point in the healthy and cancerous (metastasis) parts, respectively. The DOCP values are the average values measured at four different points in each part. The dependence of the DOCP value on the incident angle $\theta$ with $\varphi = 0°$, shown in Figure 2(a), indicates that the DOCP values from the healthy parts are larger than those from the cancerous parts with an approximately constant difference of 0.20, except for the case with $\theta = 55°$, while the DOCP value increases with increasing $\theta$ [31]. The dependence of the DOCP value on the detection angle $\varphi$ with $\theta = 0°$, shown in Figure 2(b), indicates the two characteristics similar to those of $\theta$ dependence: the increase in the DOCP value with increasing $\theta$ or $\varphi$ and a difference of approximately 0.20 between the DOCP values obtained from both parts. The former characteristics are due to the injection efficiency of CP light and irregular reflections at the surface. When the incident angle approaches the Brewster's angle between the air and the sample surface (approximately 53°, in this study), the *p*-polarization component is dominantly penetrated into the sample, but the *s*-polarization component is mostly reflected because of the difference in reflectance at the surface [9]. Therefore, the CP of light penetrating into the tissue is decreased and the scattered light has less information inside the tissue. The result shows that almost zero CP values are observed at $(\theta, \varphi) = (55°, 0°)$. Moreover, the influence of surface reflection should be considered. Some of the incident CP light penetrates into tissues and provides information about the state of the sample. The remaining part of the incident light is irregularly reflected at the rough surface of the tissues. The sign of the DOCP of the reflected light is opposite to that of the incident CP light. As the incident direction approaches the detection direction, the component due to the irregular reflection in the total detected light becomes larger, resulting in a decrease (negative increase) in the DOCP value. When the difference between the incident and detection angles becomes less than 30°, most of the detected light is reflected light, which has insufficient information. The latter characteristic of the difference in the DOCP value corresponds reasonably well with the simulation results calculated for almost the same optical configurations (Figure 2(c) in ref. [31]), which implies that the observed difference in the DOCP value is derived from the difference in particle (cell nucleus) size. The magnitude of the detected DOCP value can be influenced by extrinsic factors such as surface reflection, while the difference in the DOCP value is intrinsic and robust in the optical configuration with an angle within the appropriate range, which provides valuable information about the state of the biological tissue.

In our experiments using incident light with a wavelength of 914 nm, the DOCP value from cancerous tissues is smaller than that from healthy tissues. However, this magnitude relation is opposite to the results shown in [16], which are obtained using the incident CP light with $\lambda = 639$ nm. Light scattering is the second radiation from dipoles excited by the first irradiation on the surface of the scatterer. The degree of depolarization varies according to the distribution of dipoles, and strongly depends on the ratio of the wavelength and size of the scatterer. We have calculated the wavelength dependence of the expected DOCP value of light scattered by non-cancerous and cancerous pseudo-biological tissue in aqueous medium with dispersing scatterers having a diameter of 6.0 and 11.0 μm, respectively, by the same calculation method in ref. [31]. The calculated DOCP values show oscillating behavior derived from spherical harmonics with a variable size parameter, $x = ka$. The DOCP values from pseudo cancerous tissue ($P_{cancer}$) are larger than those from pseudo healthy tissue ($P_{health}$) at $\lambda < 680$ nm, whereas $P_{cancer} < P_{health}$ at $\lambda > 680$ nm. These calculation results can explain both experimental results.

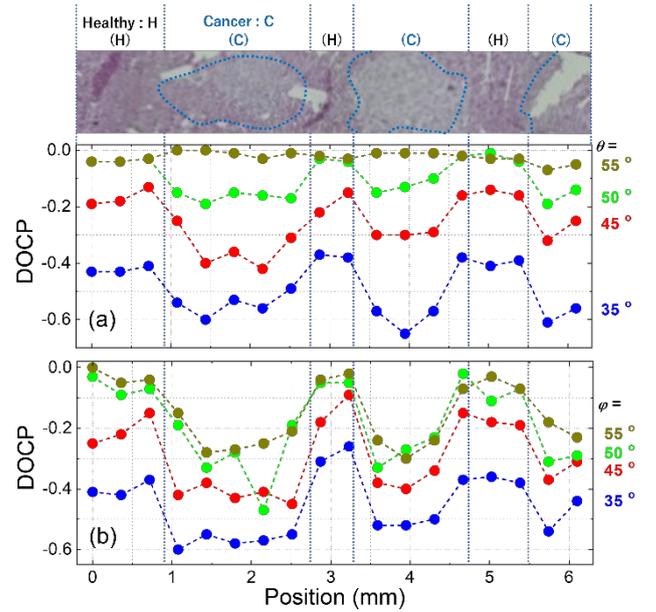

**FIGURE 3** Result of the line-scanning experiments with different (a) incident angles $\theta$ with $\varphi = 0°$ and (b) detection angles $\varphi$ with $\theta = 0°$ along the red arrow shown in Figure 1(c), respectively. The micrograph of the scanning area shown at the upper part corresponds to the probing points of the graph.

Figure 3 shows the results of the line-scanning experiments obtained at the optical configurations with $(\theta, \varphi) =$ (a) $(\theta, 0)$ and (b) $(0, \varphi)$. The data plotted in Figure 3 were obtained by acquiring one measurement at each point. The line-scanning experiments were performed at 18 points along the red arrow shown in Figure 1(b), which crosses the boundary between cancerous and healthy tissues multiple times. A micrograph of the scanning area is shown in the upper part of the graphs in which the area delineated by the blue dotted line is the cancerous parts. Except for (55°, 0°) at which the DOCP values showed almost no change, a difference of 0.1 or more in the DOCP value was observed depending on the state of the biological tissues, whether healthy or cancerous, which corresponds to the results shown in Figure 2. Distinct differences are observed in each angular configuration; however, at around the boundary between cancerous and healthy parts, steeper changes are observed with the perpendicular incidence (Figure 3 (b)) than with the oblique incidence (Figure 3 (a)). The different gradients of the DOCP value at the boundary between the oblique and perpendicular incidence could be possibly due to the different sampling (scattering) volumes and the subsequent radiation areas of the scattered light. An elongated elliptic spot due to the oblique incident beams induces the expansion of the scattering volume toward the in-plane direction inside the tissue. Accordingly, the radiation area of the scattered light is spread in the direction opposite to the incident direction. Therefore, optical configurations with oblique incidence and vertical detection have less spatial resolution. From the results of line-scanning experiments, the in-plane resolution is roughly estimated to be 0.6 and 0.3 mm for the configurations with the oblique and perpendicular incidence, respectively.

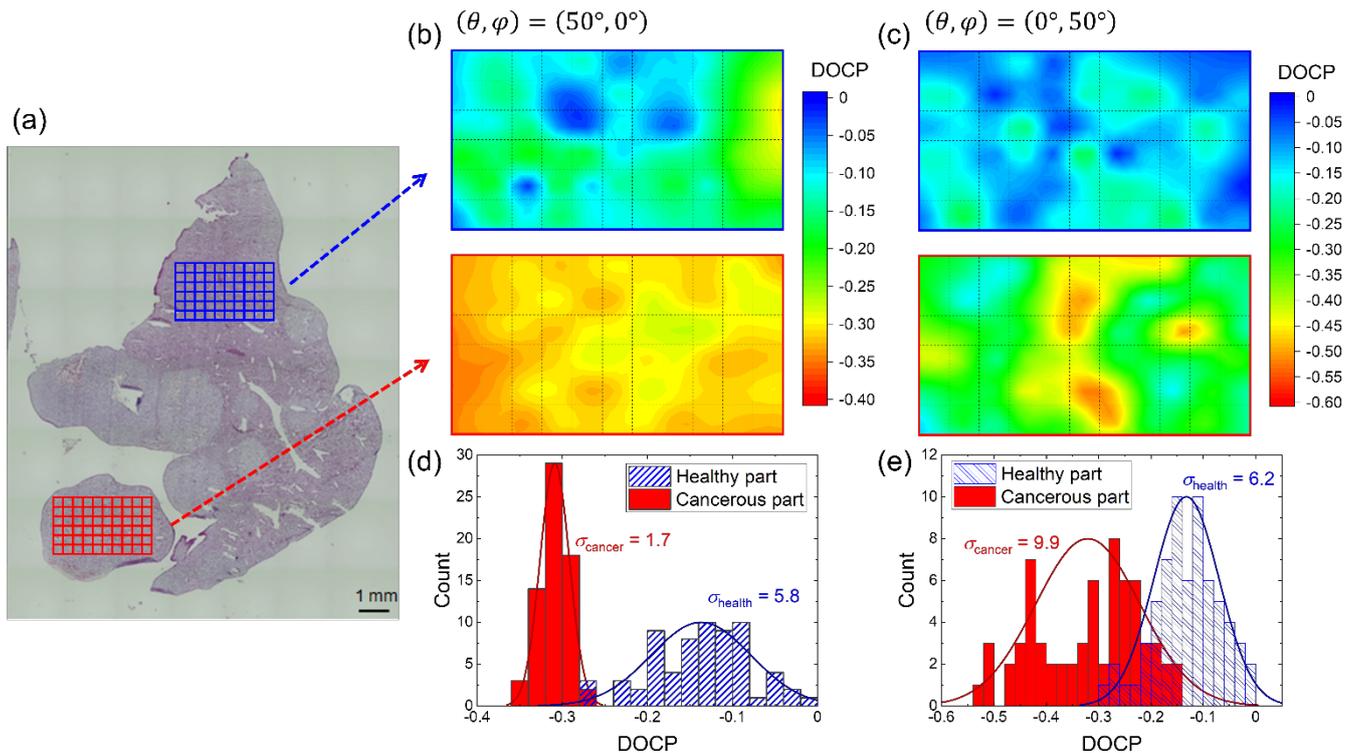

**FIGURE 3** Results of the scanning measurements. (a) The measured areas of healthy and cancerous parts, delineated by blue and red 66 squares, respectively, on the micrograph of specimen. (b)(c) The color-coded spatial distributions of the measured DOCP values in (upper panels) healthy and (lower panels) cancerous parts with $(\theta, \varphi) = $ (b) (50°, 0°) and (c) (0°, 50°), respectively. (d)(e) Histograms of the DOCP values shown in (b) and (c), respectively, together with the eye-guided distribution curves and the standard deviation values.

To evaluate the distributions of the DOCP values, scanning measurements were performed in an area that appeared as uniformly healthy and cancerous parts. The measurements were performed at each point obtained by dividing the scanning area into $6 \times 11$ lattices in length and width, which are schematically drawn on the micrograph of the specimens in Figure 4(a). The color-coded spatial distributions of the measured DOCP values at $(\theta, \varphi) = (50°, 0°)$ and $(0°, 50°)$ are shown in Figures 4 (b) and (c), respectively. The upper panels show the data in the healthy tissues, while the lower panels show the results in cancerous tissues. Figures 4(d) and (e) show histograms of the DOCP values corresponding to the data shown in Figures 4(b) and (c), respectively. At $(\theta, \varphi) = (50°, 0°)$, the dispersion of the DOCP values in the healthy part shows a wide distribution with a peak of $-0.137$ and a standard deviation of $\sigma = 5.8$, whereas the DOCP values in the cancerous part are converged at $-0.301$ with a standard deviation of $\sigma = 1.7$ (Figures 4 (b) and (d)). The almost independent dispersions provide sufficient information to distinctly identify the carcinoma lurking in healthy tissues. At $(\theta, \varphi) = (0°, 50°)$, the DOCP values are distributed with comparatively large dispersions, although the identification of cancerous tissues is possible with comparisons of data from multiple points. The peaks of the dispersions are $-0.132$ and $-0.320$, and the deviations are $\sigma = 6.2$ and 9.9 in the healthy and cancerous parts, respectively. Comparing the results of both configurations, the difference of peak in the DOCP values between healthy and cancerous parts is almost the same, whereas the dispersions are remarkably different. When the CP light is irradiated vertically, most of the penetrated light tends to progress toward the deeper layer because the forward scattering is dominant in the Mie regime. Therefore, the light experienced a large number of scattering events that were hardly discharged outward the sample, and the light scattered by a few times is dominantly detected, resulting in less accuracy for discriminating cancer.

## 4 | CONCLUSION

We experimentally investigated the applicability of scattered CP light for cancer identification in optical configurations with various angular relations between the directions of incidence and detection. An incident angle larger than the Brewster's angle for the surface of biological tissues causes a decreased penetration of polarized light into the tissue, and the small difference between the angles of incidence and detection increases surface reflection with less information. At the configuration with angles within the appropriate range, that is, $\theta \leq 53°$ and $(\theta - \varphi) \geq 30°$, the significant differences between the DOCP values obtained from the cancerous and healthy parts are observed to be approximately 0.20, which is sufficient to identify the cancer-affected area. Based on the good agreement with our previous calculations [31], here, we concluded that the difference in the DOCP values results from the different sizes of cell nuclei rather than the different reflectance values, which suggests that this technique could be applied to the identification of not only carcinomas but also other diseases accompanied by the enlargement of cell nuclei, for example, alcoholic hepatitis and ulcerous colitis. In addition to the cell nuclei, the contribution of cellular walls and other constituents also contributes to polarization scattering. These components, which are strongly associated with the anisotropic cellular shape and birefringence, can greatly contribute to polarization scattering in fibrous tissue, asymmetric complex tissue, and a tissue in which anisotropic mutation is observed. The samples used in this study consist of uniform, isotropic cell nuclei in both cancerous and healthy parts, in which the contributions of these anisotropic parameters are inconspicuous. Further research is required to investigate these contributions. In the line-scanning measurements, the obtained DOCP values change in an almost binary manner depending on the state of tissue, whether healthy or cancerous. The optical configurations with oblique incidence provide larger differences in the DOCP values, which indicates higher accuracy in identifying cancerous parts. However, the elongated elliptic spot due to oblique incidence reduces the spatial resolution and enhances the positional fluctuation. Conversely, the arrangements with perpendicular incident light have higher spatial resolution due to a narrow sampling volume but slightly less accuracy due to fewer scattering events. These arrangements should be selected according to the objective disease, organ, apparatus, and environment. Moreover, an almost constant difference in the DOCP values can be obtained at the configuration with angles within the appropriate range, which makes it permissible to incline the surface of measuring tissues against the fixed optical system. This indicates that this technique is useful even in environments where it is difficult to fix the spatial arrangement between the optics and the target, such as *in vivo* observations with an endoscope.

## ACKNOWLEDGMENTS

This work was partially supported by KAKENHI (Nos. 17K14104, 18H03878, and 19H04441) of the Japan Society for Promotion of Science (JSPS), the Cooperative Research Project of Research Center for Biomedical Engineering, Futaba Foundation, Spintronics Research Network of Japan (Spin-RNJ), and a Grant-in-Aid for Challenging Research, Organization of Fundamental Research, Tokyo Institute of Technology. The authors acknowledge technical support from the Semiconductor and MEMS Processing Division of the Technical Department of Tokyo Institute of Technology. The authors acknowledge Profs. H. Munekata and J. Yoshino for fruitful discussions and technical support at the Tokyo Institute of Technology (TIT). We would like to thank Editage (www.editage.com) for English language editing.

## DATA AVAILABILITY STATEMENT

The data that support the findings of this study are available from the corresponding author upon reasonable request.

## AUTHOR BIOGRAPHIES

Please see Supporting Information online.